## Research Article

# Profile-Based Ad Hoc Social Networking Using Wi-Fi Direct on the Top of Android

Nagender Aneja and Sapna Gambhir

*YMCA University of Science and Technology, Faridabad, India*

Correspondence should be addressed to Nagender Aneja; naneja@gmail.com

Ad hoc social networks have become popular to support novel applications related to location-based mobile services that are of great importance to users and businesses. Unlike traditional social services using a centralized server to fetch location, ad hoc social network services support infrastructure-less real-time social networking. It allows users to collaborate and share views anytime anywhere. However, current ad hoc social network applications either are not available without rooting the mobile phones or do not filter the nearby users based on common interests without a centralized server. This paper presents an architecture and implementation of social networks on commercially available mobile devices that allow broadcasting name and a limited number of keywords representing users' interests without any connection in a nearby region to facilitate matching of interests. The broadcasting region creates a digital aura and is limited by the Wi-Fi region that is around 200 meters. The application connects users to form a group based on their profile or interests using the peer-to-peer communication mode without using any centralized networking or profile-matching infrastructure. The peer-to-peer group can be used for private communication when the network is not available.

## 1. Introduction

Online social networks, for example, Facebook, LinkedIn, or Twitter, are now highly popular among people, and the trend to use the social networking applications on the mobile device is continuously increasing. The pattern of using social networking on the mobile device is being exploited by researchers and service providers to provide location-based social networking [1–3]. Examples of location-based social networking include the Facebook's feature to find exotic locations or friends nearby in a geographical region. However, current social networking applications do not provide location-based services without accessing the present site of a user, and many users consider this a privacy risk. Furthermore, limited data plans for the mobile device and high cost of international roaming constraint users to communicate even in the nearby region. Thus, the current trend is to decentralize the online social network [4].

Ad hoc social network, one-to-one or multipeer connection, can solve this problem of privacy and help facilitate the communication in a nearby region without using the centralized infrastructure. There are numerous applications; for example, it may also be useful in a business meeting where the distribution of physical business cards is not a convenient method, but the e-cards can be conveniently distributed in a peer-to-peer network to all individuals present nearby. Another application is communication among passengers in an airplane for game playing by children especially when the flight duration is long or for anonymous chatting among interested passengers or with crew members. Furthermore, ad hoc social network can help to communicate in case of natural disasters or government censorship.

Ad hoc wireless peer-to-peer technology connects devices to create a communication group for social interaction. This paper presents a Wi-Fi peer-to-peer-based mechanism called OffAT (OFFline chAT) [5] that helps to find people with similar interests in a nearby region [1, 6, 7] and allows sharing text or handwritten messages without any centralized server. The mechanism can be used to further develop applications to cater needs of business users to distribute business cards during international conferences, seminars, or meetings for networking when people may have smartphones but do not have access to the Internet.

This paper is organized as follows: Section 2 provides the review of related publications. Section 3 provides research motivation. Section 4 provides architecture and implementation comprising key components used for location-based ad hoc social networking. Section 5 provides results and analysis.

## 2. Related Work

The ad hoc social network (ASN) has the potential to connect nearby users who should be connected based on similar interests. Eagle and Pentland [8] introduced a service to introduce proximate users by doing profile matching at a central server. The service alerted users of similar interests in a nearby region. Zhang et al. [9] introduced the multihop social network to broadcast promotional offers to nearby users without using the Internet or GPS. The authors used the dynamic source routing ad hoc network protocol to deploy the multihop social network. Zhang et al. [10] considered ASN as the extension of the online social network to further increase global communication. The authors provided an architecture comprising four layers, namely, application, community, network, and device layers. The functions of profile management were implemented in the community layer. Wang et al. [11] implemented the peer-to-peer social network based on Wi-Fi Direct to facilitate communication among nearby users without a centralized infrastructure.

Some researchers have also presented middleware solutions; for example, Aneja and Gambhir [12] published a four-layer architecture with application, transport, ad hoc social, and ad hoc communication layers. Similarly, Bellavista and Giannelli [13] proposed the spontaneous multihop network using a three-layer architecture including the IP layer, spontaneous multihop layer, and semantic dispatching layer. The architecture uses a one-hop network and makes use of the dispatching layer to deliver packets to other nodes without knowing the destination. Bottazzi et al. [14] presented the socially aware and mobile architecture that provides the roaming social network and groups proximate users with similar interests.

Rahman and Hossain [15] developed nine mobile applications using a massive ad hoc social network. The applications are based on Wi-Fi, 4G, Wi-Fi Direct, and other emergency Internet access points. The mobile apps provide nearby services and tools to contact social friends based on the current context and previous history. Ilkhechi et al. [16] provided the decentralized location service scheme that can be used to identify the location in the ad hoc network. The nodes first obtain knowledge of surroundings, and the location of the target is computed based on the knowledge.

Shu et al. [17] presented Talk2Me that is a device-to-device augmented reality social network and allows people to exchange messages with nearby users. Casetti et al. [18] discussed inter- and intragroup communication technologies using the Wi-Fi Direct protocol. The number of nodes in one group that can participate in the communication depends on the IP class addressing scheme; however, Casetti et al. [18] provided a mechanism to extend the group and the range of the network by having multigroup communication. The multigroup is created by allowing a group owner to become a legacy client or a relay client in another group. However, the groups formed with clients are general groups without doing profile or interests matching. The probability of clients leaving may be high when they do not see interests being matched in the group.

Therefore, prior published works have solved issues related to profile matching at a central server and have been able to provide communication among nearby users without profile matching. As a result, there is a need for a system and a method that provide profile matching of nearby users without using the Internet or central server and allow sharing text or files with the matched users.

## 3. Research Motivation

Most portable mobile devices today are equipped with Wi-Fi Direct, Bluetooth, and Wi-Fi short-range wireless technologies which can support spontaneous on-the-go social networking by connecting users with similar interests using the ad hoc communication mode. Wi-Fi Direct or P2P technology allows devices to connect to each other to form groups. The devices negotiate roles, and one of them becomes the group owner, and the other devices connect to the owner as a client device. The devices use their MAC address as their device ID for discovery and communication and a temporary MAC address for all frames within a group.

Although direct device-to-device connectivity via the ad hoc mode was available in IEEE 802.11, it is still not widely available in the devices. Furthermore, 802.11z, Tunneled Direct Link Setup enables device-to-device communication, but devices need to be associated with the same access point. Wi-Fi P2P is based on the IEEE 802.11 infrastructure technology, but devices negotiate, and one of the devices becomes the soft access point. In other words, Wi-Fi P2P does not need a centralized fixed physical infrastructure, and any device with Wi-Fi Direct enabled can participate in the negotiation. Wi-Fi Direct or P2P specification is still at an early stage, and the many researchers have started implementing the technology.

The proposed social networking protocol and implementation, OffAT, provides social applications including interests similarity and communication between similar users that has been built using Wi-Fi Direct in the commercially available devices. OffAT allows users to broadcast user's interests in a nearby region without using any centralized infrastructure and performs interests similarity matching and calculates profile similarity to assist users in finding a person of similar interests nearby.

```
// MainActivity.java
public void startOffatAddListener () {
    startOffat.setOnClickListener ( new
        View.OnClickListener () {
    @Override
    public void onClick (View v) {
        WifiManager wifi = (WifiManager)
            getApplicationContext ().getSystemService (Context.WIFI_SERVICE);
        wifi.reconnect ();
        if (!e1.getText ().toString ().equals ("") &&
            !e2.getText ().toString ().equals ("")) {
            GlobalData.name = e1.getText ().toString ();
            GlobalData.interests =
                e2.getText ().toString ().replaceAll ("\\s+",",").replaceAll (",+",",");
            Intent intent = new Intent (v.getContext (),
                WiFiDirectActivity.class);
            startActivity (intent);
        }
    }
    } );
}
// WiFiDirectActivity.java
    public void setNameInterests () {
    try {
    Method method =
        manager.getClass ().getMethod ("setDeviceName",
    WifiP2pManager.Channel.class, String.class,
        WifiP2pManager.ActionListener.class);
    method.invoke ( manager, channel,
        GlobalData.name+"#"+GlobalData.interests,
        new WifiP2pManager.ActionListener () {
    public void onSuccess () {}
    public void onFailure (int reason) {}
    } );
    } catch (Exception e) {
    Toast.makeText (getApplicationContext (),"Unable
        send User name and
        Interests",Toast.LENGTH_SHORT).show ();
    }
    }
```

ALGORITHM 1

## 4. Architecture and Implementation

The main components of the proposed mechanism are device discovery, interest-based social network, and intergroup communication.

*4.1. Device Discovery.* Wi-Fi Direct devices discover each other using device discovery, wherein a P2P device selects a listen channel and alternates between the search state and listen state. The time for each state is allocated randomly between 100 ms and 300 ms but can be configured. Some researchers have used MAC ID as the device ID in the discovery process or some confidential ID that limits to connect with only a specified user [19] or using an Internet connection [20, 21]. For example, MobiClique [22] used Bluetooth for device discovery to locate nearby users and a central server for matching the user profiles. Profile matching at a central server is only feasible in some cases due to the additional requirement of the server or infrastructure.

The proposed method in this paper provides the mechanism and user interface to configure device ID as user name#user interests. The device ID by default is MAC address and is broadcasted as SSID in the nearby region. This paper proposes an implementation of changing device ID to user name#user interests so that interests can be broadcasted in the nearby region. The total length of name#user interests is restricted to 32 characters as per the IEEE standard for the length of SSID [23]. The broadcasted interests can be representative keywords to be used as the first layer to filter users before actual connection. The representative keywords can be the most frequent keywords that appear in users' prior actions including browsing history.

Using device ID to match user interests helps to reduce network overhead of establishing a social network of users who are not similar to each other and likely to not involve in social applications. The below code explains the methodology used to extract user ID and interests in *MainActivity.java* and set the device ID as user name and interests in *WiFiDirectActivity.java* in Java (Algorithm 1).

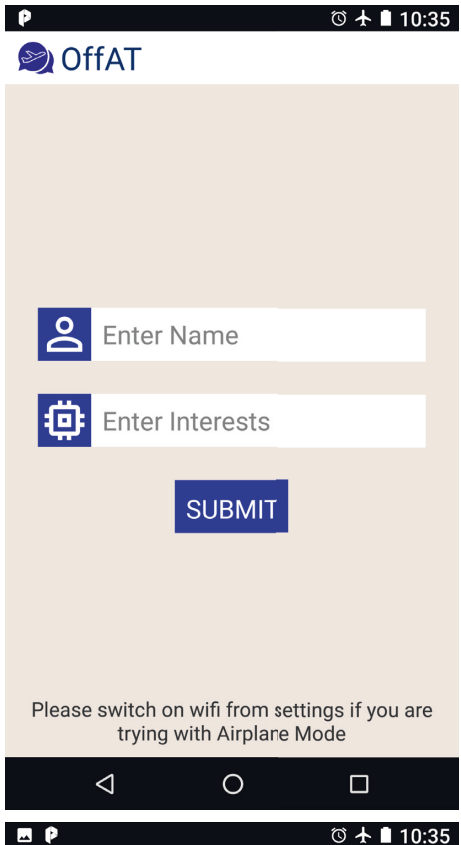

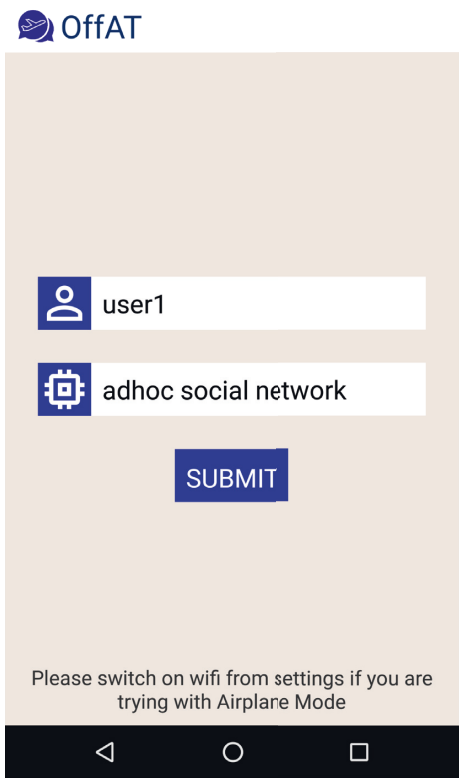

Figure 1: User interface to configure device ID.

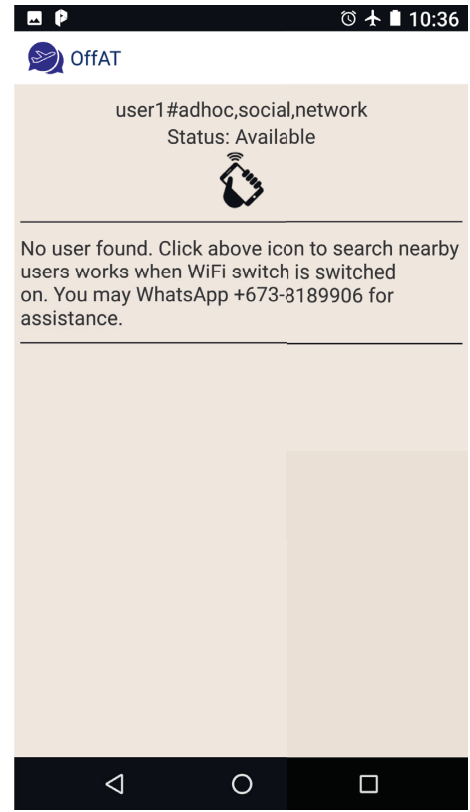

Figure 2: Device ID (displaying user name#user interests) in the discovery phase with status as available.

Figure 1 displays a user interface to configure device ID. Interests are automatically separated in the backend as a list of keywords based on space or comma operator as shown in the code. Modifying device ID to interests solves the following purposes:

(1) All user devices can access interests of other users without any centralized infrastructure
(2) Device can compute interests similarity and can decide whether to connect with other users or not
(3) Computed similarity can also be used to accept or reject any incoming connection request
(4) Reduces network overhead due to exchanging profiles

A device can have two statuses such as available or connected. A connected device can either be a group owner or an existing group member. Figure 2 shows the device discovery phase and displays the device status as available.

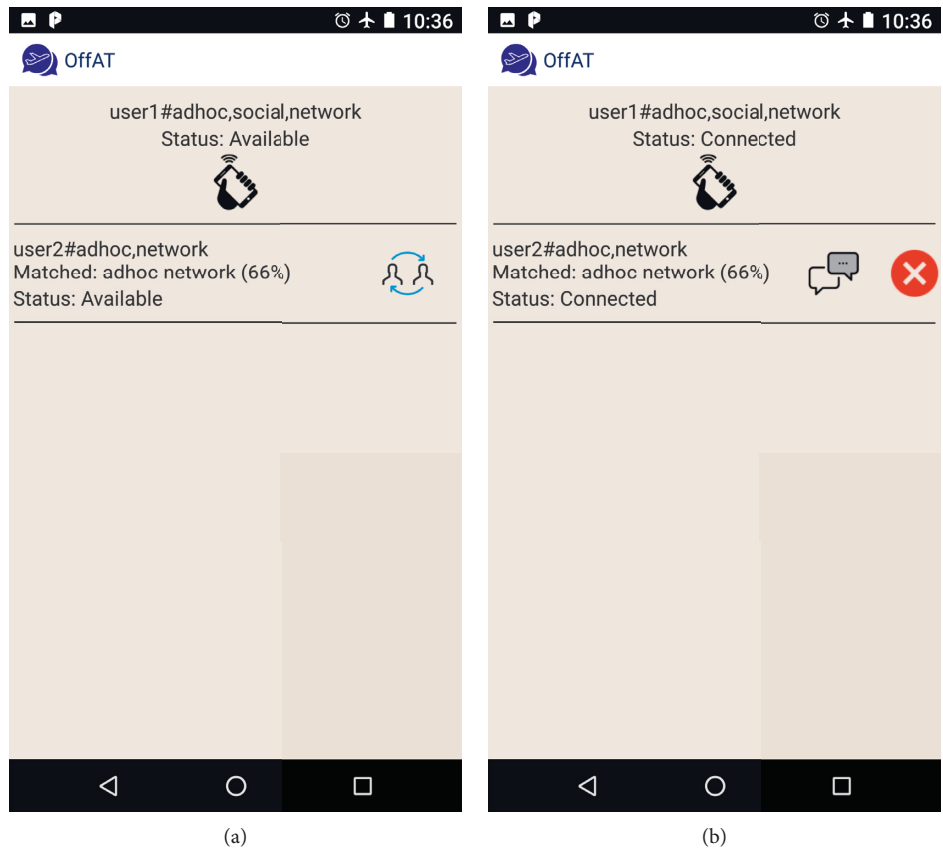

Figure 3: Profile/interests similarity with nearby devices as available/connected. (a) Nearby available devices with their interests and profile similarity. (b) Nearby connected devices with an option to chat/disconnect.

*4.2. Profile Matching.* The profile matching is the core of social networking, especially location-based social networking. Figure 3(a) displays a user device with device ID as user name#interests as available along with nearby available devices, and Figure 3(b) displays a user connected with user2. The discovery process can be manually started by clicking the hand icon available below user name and interests in case no user is found. The OffAT scans interests of nearby users and computes profile similarity based on matched keywords. The option to send a request to a neighboring user, preferably a user with high profile similarity, is displayed along the right side of each user. The list of users can quickly be scrolled to see all users if the number of users is high. Once the other user accepts the connection request, the status of both users will get connected as shown in Figure 3(b). The option to chat or disconnect and start the connection or discovery phase is displayed to the connected user as shown in Figure 3(b).

A number of researchers proposed creating the weighted profile from the user browsing history and/or prior user action [6, 24] and cosine similarity to match the profiles. The cosine similarity of two user profiles varies between 0 and 1. Cosine similarity is 1 when the angle is 0 meaning the two profiles are exactly the same. In a vector space model [24], a user profile is represented by $\{p_1, p_2, \ldots, p_n\}$, where the elements are different keywords. A union of all keywords of profiles is considered before computing cosine similarity. The weights represent the number of times a word appears in a browsing history or prior user action [6, 24]. Gambhir et al. [25] found that cosine similarity is not a correct method especially in case of a weighted profile. For the implementation of the mechanism and simplicity, the profile matching matches keywords to find the profile similarity percentage.

*4.3. Social Communication.* Once the users are connected based on their profile, they can start chatting or sharing files. This phase may also be used to add more members to the group by sending a connection request to the server designated in negotiation when the request is sent to connect. OffAT also allows exchanging handwritten notes as shown in Figure 4. Multihop communication can be facilitated by configuring a device as a client as well as a group owner where the device alternates roles based on time-sharing.

*4.4. Security.* Wireless networks are vulnerable to eavesdropping and intrinsically exposed to both active and passive security attacks. This presents a challenge in the incorporation of key-based cryptographic mechanisms for ad hoc networks because there is no trusted authority to provide certification or a centralized key distributor. Since OffAT mobile devices use Wi-Fi Direct, the security protocol implements WPA2 (Wi-Fi Protected Access II) which offers a stronger encryption algorithm known as AES (Advanced Encryption Standard). Mobile devices implement WPA2 (Wi-Fi Protected Access II) as guided by the Wi-Fi Alliance's Wi-Fi Direct certification. With stronger encryption than TKIP (Temporal Key Integrity Protocol), WPA2 also provides AES (Advanced Encryption Standard) support courtesy of the 64-digit hexadecimal keys (preshared key (PSK)). Communication privacy and security are therefore provided through WPA2 and Wi-Fi Direct.

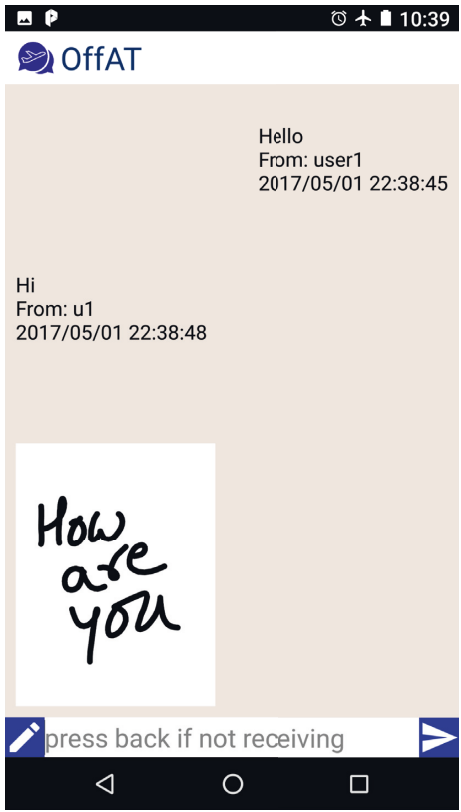

Figure 4: Exchange of text and handwritten messages in the airplane mode.

This implementation enables users to maintain privacy because the users can share user ID and limited interests without disclosing their e-mail ID or phone number unless they want to explicitly and are at liberty to decline linking requests (invitations) from other nearby users or disconnect. Furthermore, the connection requests are automatically declined after 30 seconds when the request is not responded. This preventive mechanism further protects data integrity by authenticating sources and sequencing of messages.

## 5. Results

The app, available at Google Play Store [5], is developed for the Android platform to simulate the proposed ASN. The screenshots shown in this paper are from two devices (Nexus 6P) with one device in the airplane mode and the other using the telecommunication network and also connected to a Wi-Fi router in addition to participating in the ad hoc social network. During implementation and testing, it was observed that switching on Wi-Fi is required to participate in Wi-Fi Direct communication. Therefore, there is a limitation in the airplane mode to manually switch on the Wi-Fi without even being connected to any centralized infrastructure. Although Android allows activating Wi-Fi programmatically, testing results with users indicated that this is not better since it sometimes delays the process of reconfiguring device ID. Furthermore, asking users to manually switch on Wi-Fi provides the time that an app needs to modify the device ID with user ID and interests.

The users are able to exchange text and handwritten notes among other nearby similar users without any centralized network infrastructure. Implementation results and feedback posted at Google Play Store indicated user preferences towards location-based social networking. There are few suggestions by users to fix user name and interests, and they can be configured by Shared Preferences in the Android. Future work may include to automatically exchange hello packets with nearby or connected users for displaying them as online/offline.

## Data Availability

The research paper proposes a mobile app for location-based social networking. The app is available at Google Play Store, and link is available in the references.

## Conflicts of Interest

The authors declare that they have no conflicts of interest.

## References

[1] Y. Wang, L. Wei, A. V. Vasilakos, and Q. Jin, "Device-to-device based mobile social networking in proximity (MSNP) on smartphones: framework, challenges and prototype," *Future Generation Computer Systems*, vol. 74, pp. 241–253, 2017.

[2] H. Li, K. Bok, and J. Yoo, "P2P based social network over mobile ad-hoc networks," *IEICE Transactions on Information and Systems*, vol. 97, no. 3, pp. 597–600, 2014.

[3] S. Gambhir and N. Aneja, "Ad hoc social network: a comprehensive survey," *International Journal of Scientific and Engineering Research*, vol. 4, no. 8, pp. 156–160, 2013, http://www.ijser.org/researchpaper%5CAd hoc-Social-Network-A-Comprehensive-Survey.pdf.

[4] B. Guidi, M. Conti, and L. Ricci, "P2P architectures for distributed online social networks," in *Proceedings of International Conference on High Performance Computing and Simulation (HPCS)*, pp. 678–681, IEEE, Helsinki, Finland, July 2013.

[5] N. Aneja, "OffAT-chat in aeroplane mode," 2016, http://offat.w3decode.com/.

[6] N. Aneja and S. Gambhir, "Geo-social semantic profile matching algorithm for dynamic interests in aAd hoc social network," in *Proceedings of IEEE International Conference on Computational Intelligence Communication Technology*, pp. 354–358, Ghaziabad, India, February 2015.

[7] Z. Mao, J. Ma, Y. Jiang, and B. Yao, "Performance evaluation of WiFi Direct for data dissemination in mobile social networks," in *Proceedings of Computers and*

*Communications (ISCC)*, pp. 1213–1218, IEEE, New Delhi, India, July 2017.
[8] N. Eagle and A. Pentland, "Social serendipity: mobilizing social software," *IEEE Pervasive Computing*, vol. 4, no. 2, pp. 28–34, 2005.
[9] L. Zhang, X. Ding, Z. Wan, M. Gu, and X.-Y. Li, "WiFace: a secure geosocial networking system using wifi-based multi-hop MANET," in *Proceedings of 1st ACM Workshop on Mobile Cloud Computing and Services: Social Networks and Beyond-MCS'10*, p. 3, ACM, San Francisco, CA, USA, June 2010.
[10] D. Zhang, D. Zhang, H. Xiong, C.-H. Hsu, and A. V. Vasilakos, "BASA: building mobile Ad hoc social networks on top of android," *IEEE Network*, vol. 28, no. 1, pp. 4–9, 2014.
[11] Y. Wang, A. V. Vasilakos, Q. Jin, and J. Ma, "A wi-fi direct based p2p application prototype for mobile social networking in proximity (MSNP)," in *Proceedings of 12th International Conference on Dependable, Autonomic and Secure Computing (DASC)*, pp. 283–288, IEEE, Dalian, China, June 2014.
[12] N. Aneja and S. Gambhir, "Middleware architecture for aAd hoc social network," *Research Journal of Applied Sciences, Engineering and Technology*, vol. 13, no. 9, pp. 690–695, 2016.
[13] P. Bellavista and C. Giannelli, "Middleware for semantic multicast in spontaneous multi-hop networks," in *Proceedings of International Conference on Mobile Wireless Middleware, Operating Systems, and Applications*, pp. 45–61, Springer, Berlin, Germany, November 2012.
[14] D. Bottazzi, R. Montanari, and A. Toninelli, "Context-aware middleware for anytime, anywhere social networks," *IEEE Intelligent Systems*, vol. 22, no. 5, pp. 23–32, 2007.
[15] M. A. Rahman and M. S. Hossain, "A location-based mobile crowdsensing framework supporting a massive ad hoc social network environment," *IEEE Communications Magazine*, vol. 55, no. 3, pp. 76–85, 2017.
[16] A. R. Ilkhechi, I. Korpeoglu, U. Güdükbay, and Ö. Ulusoy, "PETAL: a fully distributed location service for wireless ad hoc networks," *Journal of Network and Computer Applications*, vol. 83, pp. 1–11, 2017.
[17] J. Shu, S. Kosta, R. Zheng, and P. Hui, "Talk2Me: a framework for device–to–device augmented reality social network," in *Proceedings of International Conference on Pervasive Computing and Communications (PerCom)*, Athens, Greece, December, 2018, http://www.cse.ust.hk/~panhui/papers/talk2me.percom18.pdf.
[18] C. E. Casetti, C. F. Chiasserini, Y. Duan, P. Giaccone, and A. P. Manriquez, "Data connectivity and smart group formation in wi-fi direct multi-group networks," *IEEE Transactions on Network and Service Management*, vol. 15, no. 1, pp. 245–259, 2018.
[19] J. Joy, E. Chung, Z. Yuan, J. Li, L. Zou, and M. Gerla, "DiscoverFriends: secure social network communication in mobile ad hoc networks," *Wireless Communications and Mobile Computing*, vol. 16, no. 11, pp. 1401–1413, 2016.
[20] V. Smailovic and V. Podobnik, "Bfriend: context-aware ad hoc social networking for mobile users," in *Proceedings of 35th International Convention MIPRO*, pp. 612–617, IEEE, Piscataway, NJ, USA, May, 2012.
[21] V. Smailović and V. Podobnik, "BeFriend: a context-aware aAd hoc social networking platform," *Automatika*, vol. 57, no. 1, pp. 58–65, 2016.
[22] A.-K. Pietiläinen, E. Oliver, J. LeBrun, G. Varghese, and C. Diot, "MobiClique: middleware for mobile social networking," in *Proceedings of 2nd ACM workshop on Online Social Networks*, pp. 49–54, ACM, Barcelona, Spain, August, 2009.
[23] IEEE, *IEEE Standard for Information Technology-Telecommunications and Information Exchange between Systems-Local and Metropolitan Area Networks-Specific Requirements-Part 11: Wireless LAN Medium Access Control (MAC) and Physical Layer (PHY) Specifications*, IEEE, Piscataway, NJ, USA, 2007.
[24] J. Lee and C. S. Hong, "A mechanism for building Ad hoc social network based on user's interest," in *Proceedings of 13th Asia-Pacific Network Operations and Management Symposium*, pp. 1–4, Taipei Taiwan, September 2011.
[25] S. Gambhir, N. Aneja, and L. C. De Silva, "Piecewise maximal similarity for Ad hoc social networks," *Wireless Personal Communications*, vol. 97, no. 3, pp. 3519–3529, 2017.